\documentclass[a4paper,fleqn,usenatbib]{mnras}
\usepackage[T1]{fontenc}
\usepackage{ae,aecompl}
\usepackage{graphicx}
\usepackage{amsmath}
\usepackage{amssymb}
\usepackage{txfonts}

\def\spirou#1{{#1}}

\def\karl#1{{#1}}
\def\marvin#1{{#1}}

\newcommand{\ltsima}{$\; \buildrel < \over \sim \;$}
\newcommand{\lsim}{\lower.5ex\hbox{\ltsima}}
\newcommand{\gtsima}{$\; \buildrel > \over \sim \;$}
\newcommand{\gsim}{\lower.5ex\hbox{\gtsima}}
\newcommand{\bra}{\langle}
\newcommand{\ket}{\rangle}

\newcommand{\dd}{\mathrm{d}}

\newcommand{\chip}{{\chi^\prime}}

\newcommand{\likeli}{\mathcal{L}}

\title[Optimising tomography]
{Optimising tomography for weak gravitational lensing surveys}
\author[M. Sipp, B.M. Sch{\"a}fer, R. Reischke]
{Marvin Sipp$^1$, Bj{\"o}rn Malte Sch{\"a}fer$^1$\thanks{e-mail: bjoern.malte.schaefer@uni-heidelberg.de}, Robert Reischke$^{2,3}$\\
$^1$Astronomisches Rechen-Institut, Zentrum f{\"u}r Astronomie der Universit{\"a}t Heidelberg, Philosophenweg 12, 69120 Heidelberg, Germany\\
$^2$Department of Physics, Israel Institute of Technology, Technion, 3200003 Haifa, Israel\\
$^3$Department of Natural Sciences, The Open University of Israel, 1 University Road, P.O. Box 808, Ra'anana 4353701, Israel}

\begin{document}
\pagerange{\pageref{firstpage}--\pageref{lastpage}}
\pubyear{2020}
\maketitle
\label{firstpage}

\begin{abstract}
\karl{The subject of this paper is optimisation of weak lensing tomography: We carry out numerical minimisation of a measure of total statistical error as a function of the redshifts of the tomographic bin edges by means of a Nelder-Mead algorithm in order to optimise the sensitivity of weak lensing with respect to different optimisation targets. Working under the assumption of a Gaussian likelihood for the parameters of a $w_0w_a$CDM-model and using \textsc{Euclid}'s conservative survey specifications, we compare an equipopulated, equidistant and optimised bin setting and find that in general the equipopulated setting is very close to the optimal one, while an equidistant setting is far from optimal and also suffers from the ad hoc choice of a maximum redshift. More importantly, we find that nearly saturated information content can be gained using already few tomographic bins. This is crucial for photometric redshift surveys with large redshift errors. We consider a large range of targets for the optimisation process that can be computed from the parameter covariance (or equivalently, from the Fisher-matrix), extend these studies to information entropy measures such as the Kullback-Leibler-divergence and conclude that in many cases equipopulated binning yields results close to the optimum, which we support by analytical arguments.}
\end{abstract}


\section{Introduction}
Gravitational lensing refers to the change in shape of distant galaxies by differential gravitational deflection of light rays from these sources \citep[see e.g.][ for reviews]{2001PhR...340..291B, hoekstra_weak_2008, 2010CQGra..27w3001B}. In the regime of weak cosmic shear one observes correlations in the shapes of background galaxies that are too small to be detected in individual images, as a consequence of the weak tidal gravitational fields sourced by the large-scale structure of the Universe \citep[e.g.][]{1997ApJ...484..560J, 2000MNRAS.318..625B, 2000A&A...358...30V, huterer_weak_2010}. Because gravitational fields are agnostic to the state and the kind of matter, lensing can be used as a probe of the total matter distribution and maps out the evolution of the cosmic large-scale structure. With sensitivity towards dark energy and modified gravity through their influence on structure formation, weak cosmic shear is one of the primary science motivations for wide-field cosmological surveys, for instance the \textsc{Euclid}\footnote{\href{https://www.euclid-ec.org/}{www.euclid-ec.org/}} mission \citep{EuclidStudyReport} or the Large Synoptic Survey Telescope\footnote{\href{https://www.lsst.org/}{www.lsst.org/}} \citep[\textsc{LSST,}][]{LSST2009}.

Because lensing provides a measurement of the line-of-sight projected gravitational tidal fields, some information about the evolution is lost, concerning both the background and perturbations in the gravitational fields. This information, however, can be partly recovered by making use of the redshift information of background galaxies. The sensitivity of weak lensing measurements to virtually all parameters of a $\Lambda$CDM- or a $w_0w_a$CDM-model ({\karl{cosmological constant $\Lambda$ or varying dark energy $w_0w_a$ plus cold dark matter respectively (CDM)}}) originates from the fact that weak lensing combines geometry and structure formation: Being sensitive to fluctuations in the gravitational potential, lensing spectra are roughly proportional to $(\Omega_m\sigma_8)^2$, and as they are an integral over the CDM spectrum, there is sensitivity to the Hubble-parameter $h$ and the spectral index $n_s$. Furthermore, since lensing combines fluctuations on different scales, the background expansions enters, and as the measurable distance to the lensed galaxies is redshift $z$, all parameters in the Hubble function matter, i.e. not only the density parameters but also the dark energy equation of state parameters. Structure growth is encapsulated in the growth function, where the parameters determining the background expansion enter, but in a different weighting. 

Ideally, this leads to 3-dimensional spectra, where the line-of-sight information is diluted by the finite precision in photometry and the non-uniform galaxy-distribution \citep{2005PhRvD..72b3516C, kitching_3d_2014, grassi_detecting_2014, 2016MNRAS.459.1586Z,2018MNRAS.480.3725S, 2018PhRvD..98b3522T},  but almost all of this information can be recovered by weak lensing tomography at a fraction of the computational cost. In tomography \citep{1999ApJ...522L..21H, hu_dark_2002, jain_cross-correlation_2003, takada_cosmological_2004, huterer_nulling_2005, hannestad_measuring_2006, hollenstein_constraints_2009, kayo_information_2013, kayo_cosmological_2013, munshi_tomography_2014, 2018PhRvD..98j3507S, 2018MNRAS.480.3725S}, the source galaxy sample is split into redshift bins, where one commonly chooses bin boundaries such that every bin contains the same total number of galaxies, leading to identical levels of Poisson-noise in the ellipticity spectra estimated from every pair of identical redshift bins. Alternative ways to split the information have been investigated. For example by \citet{2012MNRAS.423.3445S} who used weighting schemes based on orthogonal polynomials, which are designed to capture information that is statistically independent from the one already obtained. While this is in principle possible, the polynomials need to be either computed for a specifically anticipated cosmology or refined in a recursive way. 

The question that we would like to answer is whether non-uniform (irrespective whether in redshifts or number density), optimised binning choices can improve the statistical precision of weak lensing and yield tighter constraints on cosmological parameters. Technically, we formulate a measure of total error as a function of the bin boundaries in redshift, and numerically optimise these redshifts to give the smallest possible error. Even for the case of a Gaussian-approximated likelihood, where the entire information about the statistical error is contained in the Fisher-matrix $F_{\mu\nu}$, which corresponds to the inverse parameter covariance, $\boldsymbol{C} = \boldsymbol{F}^{-1}$, a range of possible measures of total statistical error can be {\spirou{physically}} motivated and immediately formulated: Those would include the trace $\mathrm{tr}(C) = \sum_\mu C_{\mu\mu}$, the Frobenius-norm $\mathrm{tr}(\boldsymbol{C}^2) = \sum_{\mu\nu} C_{\mu\nu}F_{\mu\nu}$, or the determinant $\mathrm{det}(\boldsymbol{C})$. In parallel, we will consider optimisations with a {\spirou{more cosmological}} motivation, for instance the dark energy figure of merit. Ultimately, we will formulate as well measures of degeneracy breaking and of the loss of information entropy (which can be expressed analytically for a Gaussian-approximated likelihood) as a target for optimisation of tomographic bins. The specifics of the optimised binning depend on the chosen target function for the optimisation. Complementary to our approach, the possibility of a binning in colour-space was investigated in \citet{2019arXiv190106495K} using the dark energy figure of merit as a metric to which we will compare {\spirou{our results}} in the discussion.

The fiducial cosmological models are spatially flat $\Lambda$CDM- or $w_0w_a$CDM-cosmologies \citep{2018arXiv180706209P}, with specific parameter choices $\marvin{\Omega_m = 0.32}$, $\marvin{n_s = 0.96}$, $\marvin{\sigma_8 = 0.816}$ and $\marvin{h=0.67}$, motivated by the Planck-observation of the cosmic microwave background. Generic quintessence models are chosen such that their equation of state is parameterised by a linear time-evolution in $w_0$ and $w_a$, in contrast to the fiducial cosmology, where $w_0=-1$ and $w_a=0$. After a summary of cosmology and weak gravitational lensing in Sect.~\ref{sect_cosmology} we outline the optimisation procedure in Sects.~\ref{sect_statistics} and~\ref{sect_optimisation}. We present the results in Sect.~\ref{sect_results}, followed by a summary in Sect.~\ref{sect_summary}.

\section{cosmology and cosmic shear basics}\label{sect_cosmology}
Under the symmetry assumptions of Friedmann-Lema{\^i}tre-cosmologies all fluids are entirely characterised by their density and their equation of state: In spatially flat cosmologies with the matter density parameter $\Omega_m$ and the corresponding dark energy density $1-\Omega_m$ one obtains for the Hubble function $H(a)=\dot{a}/a$ the expression,
\begin{equation}
\frac{H^2(a)}{H_0^2} = \frac{\Omega_m}{a^{3}} + \frac{1-\Omega_m}{a^{3(1+w_0+w_a)}}\exp[3w_a(a-1)]\;,
\end{equation}
where a linearly evolving, CPL-parameterised equation of state function $w(a)$ \citep{2001IJMPD..10..213C, 2006APh....26..102L, 2008GReGr..40..329L},
\begin{equation}
w(a) = w_0 + (1-a)w_a\;,
\end{equation}
was assumed for the dark energy fluid. The comoving distance $\chi$ is related to the scale factor $a$ through
\begin{equation}
\chi = -c\int_1^a\:\frac{\dd a}{a^2 H(a)}\;,
\end{equation}
where the Hubble distance $\chi_H=c/H_0$ sets the scale for cosmological distance measures. Cosmic deceleration $q=-\ddot{a}a/\dot{a}^2$ is related to the logarithmic derivative of the Hubble function, $2-q = 3+\dd\ln H/\dd\ln a$.

Small fluctuations $\delta$ in the distribution of dark matter grow, as long as they are in the linear regime $\left|\delta\right|\ll 1$, according to the growth function $D_+(a)$ \citep{2003MNRAS.346..573L,1998ApJ...508..483W},
\begin{equation}
\frac{\dd^2}{\dd a^2}D_+(a) +
\frac{2-q}{a}\frac{\dd}{\dd a}D_+(a) -
\frac{3}{2a^2}\Omega_m(a) D_+(a) = 0\;,
\label{eqn_growth}
\end{equation}
and their statistics is characterised by the spectrum $\bra \delta(\bmath{k})\delta^*(\bmath{k}^\prime)\ket = (2\pi)^3\delta_D(\bmath{k}-\bmath{k}^\prime)P_\delta(k)$. Inflation generates a spectrum of the form $P_\delta(k)\propto k^{n_s}T^2(k)$ with the transfer function $T(k)$. As our primary interest is a proof of principle, the accuracy of the transfer function is not very important, so we will use one which is parameterised in a straightforward way \citep{eisenstein_power_1999, eisenstein_baryonic_1998}. The spectrum $P(k)$ is normalised ot the variance $\sigma_8^2$ on the scale $R = 8~\mathrm{Mpc}/h$,
\begin{equation}
\sigma_8^2 = \int_0^\infty\frac{k^2\dd k}{2\pi^2}\: W^2(kR)\:P_\delta(k)\;,
\end{equation}
with a Fourier-transformed spherical top-hat $W(kR) = 3j_1(kR)/(kR)$ as the filter function. From the CDM-spectrum of the density perturbations the spectrum of the dimensionless Newtonian gravitational potential $\Phi$ can be obtained,
\begin{equation}
P_\Phi(k) \propto \left(\frac{3\Omega_m}{2\chi_H^2}\right)^2\:k^{n_s-4}\:T(k)^2,
\end{equation}
by applying the comoving Poisson-equation $\Delta\Phi = 3\Omega_m/(2\chi_H^2)\delta$ for deriving the gravitational potential $\Phi$ from the density $\delta$. Nonlinear structures increase the variance on small scales, which is described through a parameterisation of $P_\delta(k,a)$ proposed by \citet{2003MNRAS.341.1311S}.

In weak gravitational lensing one investigates the action of gravitational tidal fields on the shape of distant galaxies by the distortion of light bundles \citep[for reviews, please refer to][]{2001PhR...340..291B, hoekstra_weak_2008, huterer_weak_2010, 2010CQGra..27w3001B}. The lensing potential $\psi_i$ is given by a projection integral of the gravitational potential $\Phi$,
\begin{equation}
\psi_i = \int_0^{\chi_H}\dd\chi\:W_i(\chi)\Phi,
\label{eqn_lensing_potential}
\end{equation}
related through the tomographic weighting function $W_i(\chi)$,
\begin{equation}
W_i(\chi) = 2\frac{D_+(a)}{a}\frac{G_i(\chi)}{\chi}.
\end{equation}
As a line of sight-integrated quantity, the projected potential contains less information than the sourcing field $\Phi$. In order to partially regain that information, one commonly subdivides the sample of lensed galaxies into $n_\mathrm{bin}$ redshift bins and estimates the lensing spectrum for every bin combination separately. Therefore, one defines the tomographic lensing efficiency function $G_i(\chi)$,
\begin{equation}
G_i(\chi) = \frac{1}{f_i}\int^{\chi_{i+1}}_{\max(\chi,\chi_i)}\dd\chip\:
p(\chip) \frac{\dd z}{\dd\chip}\left(1-\frac{\chi}{\chip}\right)\;,
\end{equation}
with $\dd z/\dd\chip = H(\chip) / c$ and the bin edges $\chi_i$ and $\chi_{i+1}$, respectively. $f_i$ denotes the fraction of galaxies in the $i$-th bin,
\begin{equation}
	f_i = \int_{\chi_i}^{\chi_{i+1}} \dd\chip\:
	p(\chip) \frac{\dd z}{\dd\chip}\;.
\end{equation}
Commonly, weak lensing  forecasts use the parameterisation of the redshift distribution $p(z)\dd z$,
\begin{equation}
p(z)\propto \left(\frac{z}{z_0}\right)^2\exp\left[-\left(\frac{z}{z_0}\right)^\beta\right]\;.
\end{equation}
Combining all results one obtains the angular spectra $C_{\psi,ij}(\ell)$ of the tomographic weak lensing potential $\psi_i$ in the flat-sky approximation \citep{1954ApJ...119..655L},
\begin{equation}
C_{\psi,ij}(\ell) = \int_0^{\chi_H}\frac{\dd\chi}{\chi^2}\:W_i(\chi)W_j(\chi)\:P_\Phi((\ell + 1/2)/\chi,\chi)\;.
\end{equation}
{\karl{
Weak lensing convergence or shear are related to the lensing potential by applying two angular derivatives, therefore their spectra are equal to $\ell^4C_{\psi,ij}(\ell)/4$ for a flat sky. In our code, we directly use the observable convergence spectrum by relating the potential fluctuations to the matter fluctuations via the Poisson equation.}} The spectra $C_{\psi,ij}(\ell)$ are different from zero for $i\neq j$ leading to a non-diagonal covariance matrix in the construction of the Fisher-matrix. Traditionally, the choice of bin edges is such that each bin contains an identical fraction of the total number $4\pi\bar{n}$ of galaxies, but in our case the number of galaxies is variable resulting in a non-uniform shape noise term, which would nevertheless still be only present in the diagonal of the covariance matrix due to non-overlapping bins, in contrast to \citet{2012MNRAS.423.3445S}. For a standard binning with constant shape noise contribution $\sigma_\epsilon^2 n_\mathrm{bin}/\bar{n}$ one would obtain
\begin{equation}
\hat{C}_{\psi,ij}(\ell) = C_{\psi,ij}(\ell) + \frac{\sigma_\epsilon^2}{\ell^4}\frac{n_\mathrm{bin}}{\bar{n}}\delta_{ij}\;,
\end{equation}
while introducing a non-standard binning would lead to
\begin{equation}
\hat{C}_{\psi,ij}(\ell) = C_{\psi,ij}(\ell) + \frac{\sigma_\epsilon^2}{\ell^4}\frac{1}{\bar{n}f_i} \delta_{ij}\;.
\end{equation}
{\spirou{Here, we point out that our non-uniform binning schemes define non-overlapping redshift intervals, such that the Poisson noise term is in all cases diagonal, scales inversely with the fraction $f_i$ of the galaxy population and only appears in the estimation of spectra with identical bin indices, $i=j$.}}

Specifically, we compute our optimisations for the \textsc{Euclid} weak lensing survey \citep{2008arXiv0802.2522R, EuclidStudyReport} with the choices: $(i)$ a median redshift of $0.9$ and $\beta =1.5$, $(ii)$ a yield of $\bar{n}=4.7\times10^8$ galaxies per unit solid angle, $(iii)$ a sky fraction of $f_\mathrm{sky} \simeq 0.3$ and $(iv)$ a Gaussian shape noise with variance $\sigma_\epsilon^2=0.09$.

\section{statistics}\label{sect_statistics}
We compute a Gaussian-approximated likelihood for a fixed fiducial model, such that the entire information about the likelihood is contained in the Fisher-matrix. Constraints on the $\Lambda$CDM- or $w_0w_a$CDM-parameters are derived from the set of $\ell,m$-modes of the tomographic shear field $\gamma_i(\theta,\phi)$, or equivalently, the weak lensing potential $\psi_i(\theta,\phi)$.

Under the assumption of statistical isotropy of the weak lensing sky, full sky tomographic weak lensing surveys provide a measurement of $2\ell+1$ statistically independent modes for each multipole $\ell$. Constraints on cosmological parameters can be derived \citep{1997ApJ...480...22T, 2002PhRvD..66h3515H} from the set of modes $\psi_{\ell m,i}$ that are isolated for each tomography bin by spherical harmonic transform,
\begin{equation}
\psi_{\ell m,i} = \int\dd\Omega\: \psi_i(\bmath\theta)Y_{\ell m}^*(\bmath\theta).
\end{equation}
The likelihood $\likeli$ that a model $\hat{C}_\psi(\ell)$ is able to reproduce the set $\left\{\psi_{\ell m,i}\right\}$ of observed modes $\psi_{\ell m,i}$ separates in $\ell$ and $m$ according to
\begin{equation}
\likeli\left(\left\{\psi_{\ell m,i}\right\}\right) = \prod_\ell \likeli\left(\psi_{\ell m,i}|\hat{C}_{\psi,ij}(\ell)\right)^{2\ell + 1},
\end{equation}
because of the symmetry assumptions, while there is no separation in the tomographic bin index $i$,
\begin{equation}
\likeli\left(\psi_{\ell m,i}\right) =
\frac{1}{\sqrt{(2\pi)^{n_\mathrm{bin}}\mathrm{det}\hat{C}_\psi(\ell)}}\exp\left(-\frac{1}{2}\psi_{\ell m,i}(\hat{C}_{\psi}(\ell)^{-1})_{ij}\psi_{\ell m,j}\right)\;,
\end{equation}
from the fact that both the cosmic structure as well as the noise are statistically isotropic and homogeneous Gaussian random fields. This assumption is only valid in linear structure formation, where all Fourier-modes evolve in a statistically independent way. Line of sight-integrations are able to reduce the amount of non-Gaussianity in the lensing observables as a consequence of the central limit theorem, but residual non-Gaussian covariances lead to misestimations of parameters, as shown by \citet{scoccimarro_power_1999, kayo_information_2013}. We incorporate nonlinear structures effectively by increasing the variance of the fields without accounting for the deviation from Gaussianity. {\karl{For the scales we are probing a Gaussian covariance is a good approximation \citep{euclid_preparation_forecast_2019}, so that we do not include trispectrum corrections.}}

The negative logarithmic likelihood $L = -\ln\likeli$ is given by
\begin{equation}
L = \sum_\ell\frac{2\ell+1}{2}\:\left(\ln\hat{C}_{\psi,ii} + (\hat{C}_\psi^{-1})_{ij}\:\psi_{\ell m,i}\psi_{\ell m,j}\right)\;,
\end{equation}
up to an additive constant. From the data-averaged curvature $F_{\mu\nu} = \bra\partial_\mu\partial_\nu L\ket$ of the negative logarithmic likelihood one derives the Fisher matrix $F_{\mu\nu}$,
\begin{equation}
F_{\mu\nu} = \sum_\ell\frac{2\ell+1}{2}\left(
\frac{\partial}{\partial x_\mu}\ln\hat{C}_{\psi,ij}(\ell)\:\frac{\partial}{\partial x_\nu}\ln\hat{C}_{\psi,ji}(\ell)\right)\;,
\label{eqn_fisher}
\end{equation}
with $\partial/\partial x_\mu$ being derivatives with respect to individual cosmological parameters $x_\mu$. With the Fisher-matrix $F_{\mu\nu}$, or equivalently with the parameter covariance $\boldsymbol{C} = \boldsymbol{F}^{-1}$, the posterior distribution $p(x_\mu)\dd x_\mu$ for a trivial prior can be formulated as
\begin{equation}
p(x_\mu) = \sqrt{\frac{\mathrm{det}(\boldsymbol{F})}{(2\pi)^n}}\exp\left(-\frac{1}{2}\sum_{\mu\nu}(x_\mu - x^\mathrm{fid}_\mu)F_{\mu\nu}(x_\nu - x^\mathrm{fid}_\nu)\right)\;.
\end{equation}

{\spirou{Typically, we compute Fisher-matrices by summation from $\ell_\mathrm{min} = 10$ to $\ell_\mathrm{max} = 1500$ which corresponds to a conservative choice for computing \textsc{Euclid} forecasts \citep{euclid_preparation_forecast_2019}.}} The sky coverage is set to $f_\mathrm{sky} = 1/3$. {\spirou{It should be noted that we neglect correlations introduced by \spirou{non-uniform sky coverage} such that a different sky fraction just constitutes an overall prefactor as a Poissonian scaling of the errros, and does not influence the optimisation itself.}}

\section{tomography optimisation}\label{sect_optimisation}
We optimise weak lensing tomography by formulating a measure of total error as a function of the bin boundaries $z_i$ in terms of redshift, for a fixed total number of bins $n_\mathrm{bin}$. The optimal binning is found numerically through a Nelder-Mead-simplex {\spirou{(\textsc{AMOEBA})}} optimisation in the space spanned by the set of redshift bin boundaries 
\begin{equation}
\{z_i\}\coloneqq \left\{\;z_i\in\mathbb{R}_{>0}\; \big|\; z_i< z_j,\;  i,j\in [1,n_\mathrm{bin}]\; |\; i<j\;\right\}\;,
\end{equation}
 restricted with the condition that the bins are non-overlapping and ordered in redshift. Testing the Nelder-Mead-simplex optimisation for 2-bin tomography showed a fast convergence already for standard settings, and a viable initial setting in all cases are standard tomography bins with equal fractions of the total galaxy number. {\spirou{In addition, we vary the initial conditions for optimisation and make sure that Nelder-Mead algorithm always converges to wards the same point irrespective of the initial binning, see appendix~\ref{app:covergence}.}}

As a target for optimisation we use different possible measures of the total statistical error which can be derived from the Fisher-matrix $F_{\mu\nu}$ or, with the same motivation, from the parameter covariance $\boldsymbol{C} = \boldsymbol{F}^{-1}$ in a Gaussian approximation {\karl{of the posterior}}. This is equivalent to maximising the information gain. It is worthwhile to state at this point that trace invariants of $\boldsymbol{C}$ and $\boldsymbol{F}$ do in general not yield equivalent targets for optimisation even in the case of diagonal matrices, because arithmetic and harmonic means are different except if all errors are equal. Of course, only for a Gaussian distribution the covariance $C_{\mu\nu}$ follows directly from the inverse Fisher-matrix, $C_{\mu\nu} = \int\dd x_\mu\dd x_\nu\: x_\mu x_\nu p(x_\mu,x_\nu) = (F^{-1})_{\mu\nu}$, and we will briefly discuss realistic, non-Gaussian likelihoods at the end of this paper.
\begin{itemize}
\item[]We start by considering the trace of the inverse Fisher-matrix,
\begin{equation}
\mathrm{tr}(\boldsymbol{C}) = \sum_\mu C_{\mu\mu} = \sum_\mu\sigma^2_\mu,
\end{equation}
which is naturally invariant under reparameterisation and reflects the total uncertainty without respect to degeneracy or correlation. Extracting individual errors $\sigma_{\mathrm{m},\mu}^2 = C_{\mu\mu} = (F^{-1})_{\mu\mu}$ from the Fisher matrix corresponds to marginalisation, so $\mathrm{tr}(\boldsymbol{C})$ is equal to the sum of the marginalised variances. We will in parallel compute optimised bin configurations for minimising errors $\sigma_\mu$ on single cosmological parameters $x_\mu$. Conversely, extracting errors $\sigma_{\mathrm{c},\mu}^2 = 1/F_{\mu\mu}$ would correspond to conditionalisation, with both operations being equal for a one-dimensional parameter space.

\item[]{The Frobenius-norm of the inverse Fisher-matrix is given by
\begin{equation}
\mathrm{tr}(\boldsymbol{C}^2) = \sum_{\mu\nu}C_{\mu\nu}C_{\mu\nu} = \sum_{\mu\nu} r_{\mu\nu}^2\sigma_\mu^2\sigma_\nu^2\;,
\end{equation}
with the Pearson correlation coefficient $r_{\mu\nu}$, {\spirou{defined through}} $C_{\mu\nu} = r_{\mu\nu}\sigma_\mu\sigma_\nu$. The squared trace therefore carries information about degeneracies between the parameters and is likewise invariant under reparameterisation. For the uncorrelated case, $r_{\mu\nu} = \delta_{\mu\nu}$, such that the target for optimisation is $\sum_\mu\sigma_\mu^4$, giving a stronger weight to reducing large errors in the budget, as opposed to $\sum_\mu\sigma^2_\mu$ as the previous case. 
}

\item[]{The determinant of the inverse Fisher-matrix,
\begin{equation}
\mathrm{det}(\boldsymbol{C}) = \frac{1}{\mathrm{det}(F)} = \prod_\mu\sigma^2_\mu\;,
\end{equation}
measures the volume of the region bounded by the $1\sigma$-contour in parameter space, and would be equal to the product of the individual variances for a diagonal covariance, {\spirou{as indicated by the last equality}}. Again, it considers degeneracies between the parameters and is related to the logarithmic trace of $\boldsymbol{C}$, $\ln\det \boldsymbol{C} = \mathrm{tr}\ln \boldsymbol{C}$. Expressing $\ln\det \boldsymbol{C}$ for a diagonal covariance matrix in terms of the individual variances and using $\ln\det \boldsymbol{C} = \mathrm{tr}\ln \boldsymbol{C}$ leads to
\begin{equation}
\ln\det \boldsymbol{C} = \sum_\mu \ln\sigma^2_\mu\;,
\end{equation}
which shows a lesser down-weighting of small errors compared to $\mathrm{tr}(\boldsymbol{C})$ or $\mathrm{tr}(\boldsymbol{C}^2)$. Pictorially, $\ln\det \boldsymbol{C}$ is the logarithmic volume of the parameter space bounded by the $1\sigma$-probability contour.}
\item[]{Closely related to the last measure is the dark energy figure of merit, which corresponds to the volume of the ellipsoid in $(w_0,w_a)$-space, bounded by the $1\sigma$-contour:
\begin{equation}
\mathrm{FoM} = \frac{1}{\sqrt{\det \boldsymbol{C}(w_0,w_a)}}\;,
\end{equation}
after all parameters except $w_0$ and $w_a$ have been marginalised out. A larger $\mathrm{FoM}$ implies better and more precise measurements of the dark energy equation of state and its time evolution.
}
\item[]{Lastly, we consider the information entropy difference between, for instance, a CMB-prior $p_\mathrm{CMB}(x_\mu)$ and the combined measurement $p(x_\mu)p_\mathrm{CMB}(p_\mu)$ consisting of weak lensing and the CMB, in the form of the Kullback-Leibler divergence $D_\mathrm{KL}$,
\begin{equation}
D_\mathrm{KL} = \int\dd^n x\: p(x_\mu)\ln \Bigg( \frac{p(x_\mu)}{p(x_\mu)p_\mathrm{CMB}(x_\mu)}\Bigg)\;,
\end{equation}
which can, in the Gaussian approximation, be entirely expressed in terms of traces and determinants of the two involved covariance matrices. The Kullback-Leibler-divergence corresponds to the reduction in information entropy from the CMB-prior to a combined measurement and describes, very loosely, the gain in knowledge on the parameter set. Initially, we will use a Gaussian CMB-prior with a corresponding Fisher-matrix $F^\mathrm{CMB}_{\mu\nu}$ with proper marginalisation over the optical depth and the baryon-density, to make it compatible with the Fisher-matrix from weak gravitational lensing.
}
\end{itemize}

Taking logarithms of these measures is sometimes convenient from numerical points of view, but would not affect the minimisation process due to monotonicity of the logarithm. All measures are ultimately computed from the Fisher-matrix $F_{\mu\nu}$ or its inverse, the parameter covariance $C_{\mu\nu}$ and thus rely on the assumption of Gaussianity, which might not be given in the analysis of a data set.

\section{results}\label{sect_results}
{\karl{We will investigate a range of different measures as possible optimisation targets. All results are compared to the case of equidistant and equipopulated redshift bins. Although we would like to emphasise that the specific outcome in terms of bin boundaries depends on the chosen optimisation, there seems to be the general pattern that the equipopulated is very close to the optimal one. This cannot be said about the equidistant case, which performs always worse than the other two cases. In general, equidistant tomographic bins are conceptually constructed very differently from equipopulated bins and rely on the choice of a maximum redshift without taking into account the distribution of sources. The latter can of course be put in by hand, but it does not appear naturally. The advantage of this choice might be that one can choose the width of the bins such that there is a very small risk of photometric redshift outliers. For equipopulated bins the bin width around the peak of the source redshift distribution can be come quite narrow. However, given the typical scaling of redshift errors $\sigma_z \propto (1+z)$ with a constant proportionality of the order 0.01 and given the typical width of photometric bins, this seems well within reach as we will discuss later in this section.}}

\subsection{Optimisation of individual errors}
\begin{figure}
\resizebox{0.98\hsize}{!}{\includegraphics{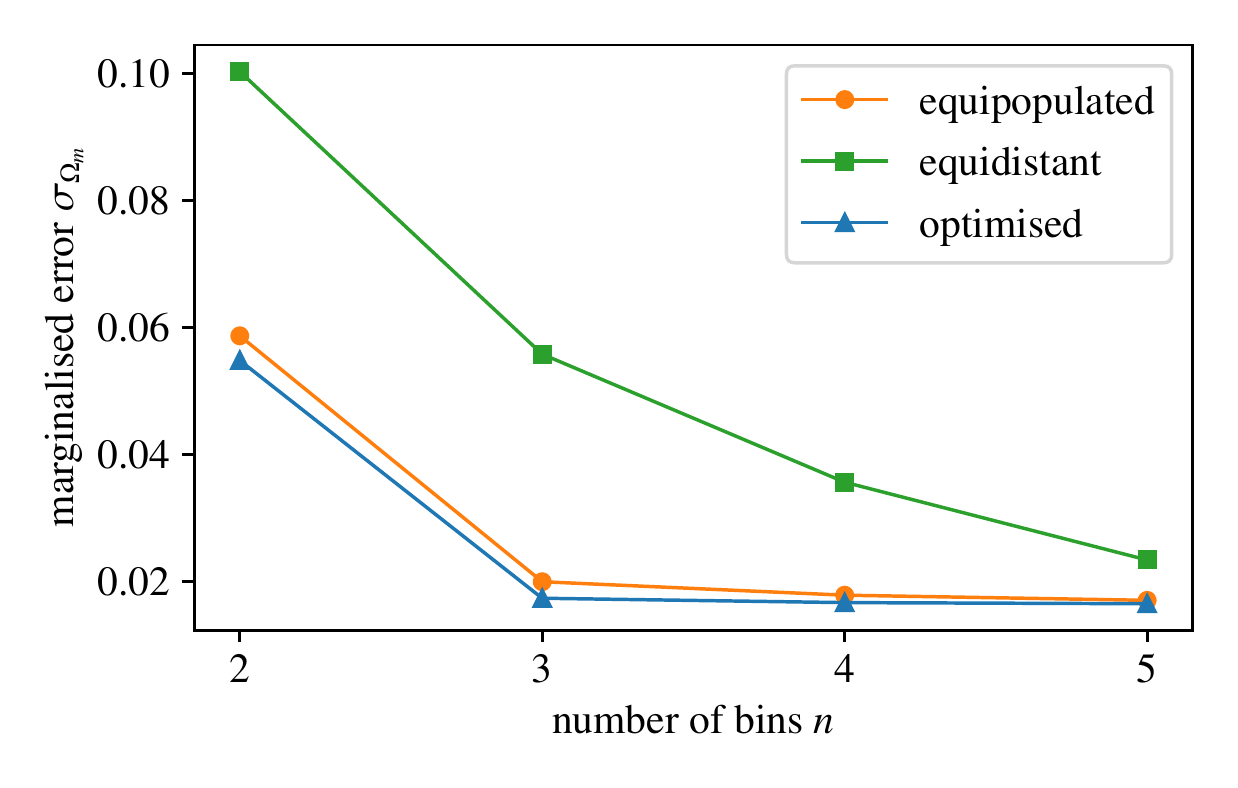}}
\caption{\karl{Marginalised error $\sigma_{\Omega_m}$ on the matter density $\Omega_m$ as a function of the number of tomographic bins, for equipopulated binning with identical fractions of the galaxy sample (orange), for a binning equidistant in redshift (green) and for a binning optimised to yield the smallest error (blue).}}
\label{omega_m_error}
\end{figure}

{\karl{As a first application we demonstrate the optimisation of individual errors, although marginalised, by suitable tomographic binning, and show the results for the matter density parameter $\Omega_m$:  The marginalised error $\sigma_{\mathrm{m}}$ as a function of the number of bins, is shown in Fig.~\ref{omega_m_error}. A significant improvement across all considered numbers of bins could be achieved with respect to the equidistant case. Interestingly, the information gain almost saturates at three tomographic bins. Finer subdivision of the sample only leads to marginally better results. It should be noted that this is paramount for photometric surveys, since the accurate propagation of photometric redshift errors is very difficult and therefore few as well as broad bins are very desirable. Equipopulated binning is found to be almost optimal in this case, but further tomographic subdivision can in fact be beneficial in breaking degeneracies in other experiments. This can be explained by the fact that the main driver of the sensitivity to $\Omega_m$ is the amplitude of the lensing power spectrum. We show in Sect.~\ref{sect_s2n} that an amplitude-like parameter is optimally constrained with equipopulated bins.}}

\begin{figure}
\resizebox{0.98\hsize}{!}{\includegraphics{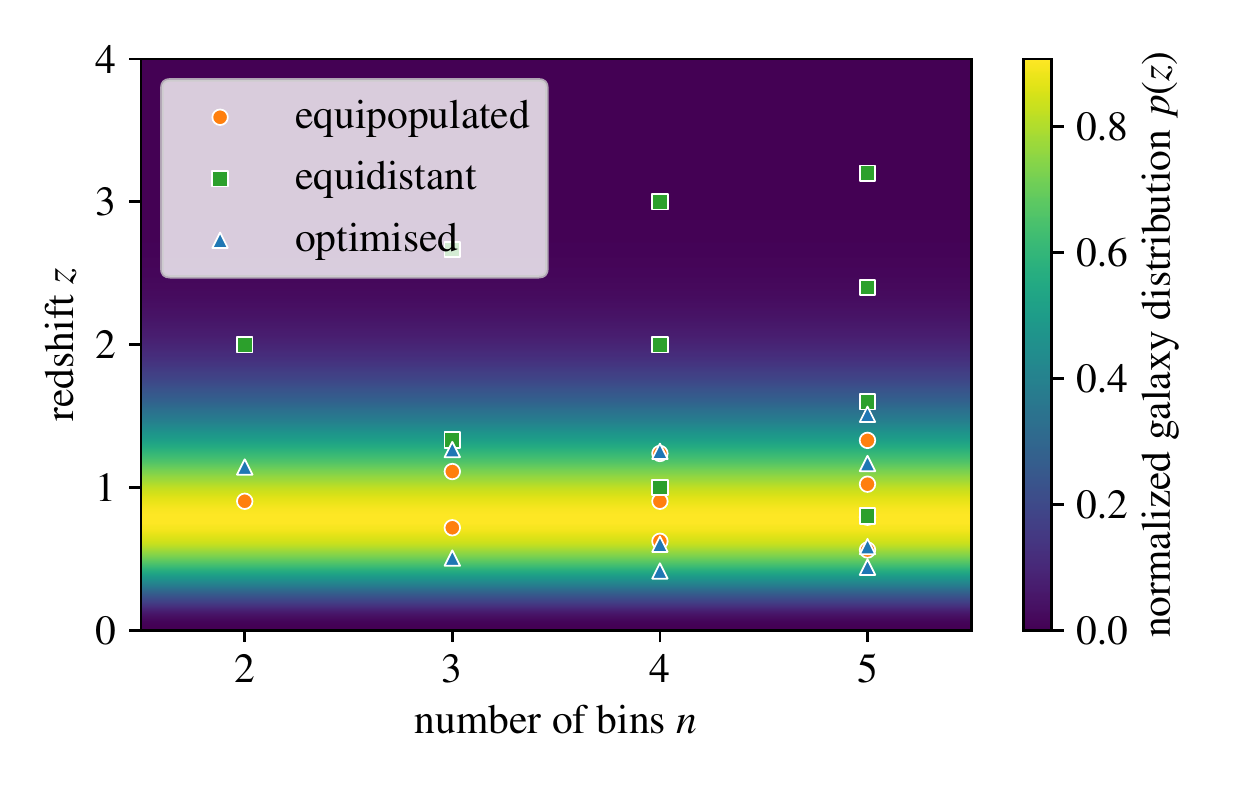}}
\caption{\karl{Galaxy distribution $p(z)\dd z$ (as the background shading) with bin edges in equipopulated binning (orange circles), equidistant bins (green squares) and with optimised (with respect to the error in $\Omega_m$) bin edges (blue triangles).}}
\label{omega_m_binning}
\end{figure}

{\karl{Fig.~\ref{omega_m_binning} illustrates the bin placement for the three cases with the galaxy distribution as a color scale. The bins for the equipopulated case almost agree with the optimal ones. Equidistant bins are usually placed at higher redshifts. The reason for this is of course the choice of the maximum redshift ($z_\mathrm{max} = 4$ in our case). If we would chose the maximum redshift between two and three the equidistant case would perform better than now. However, we would argue that equidistant bins are not a good choice in general, since it adds an additional free parameter to the problem: the maximum redshift or alternatively the bin width. Qualitatively very similar results have been obtained using $\mathrm{tr}(\boldsymbol{C})$ and $\mathrm{tr}(\boldsymbol{C}^2)$ as optimisation targets.}}

\subsection{Optimisation of the dark energy figure of merit}
\begin{figure}
	\resizebox{0.98\hsize}{!}{\includegraphics{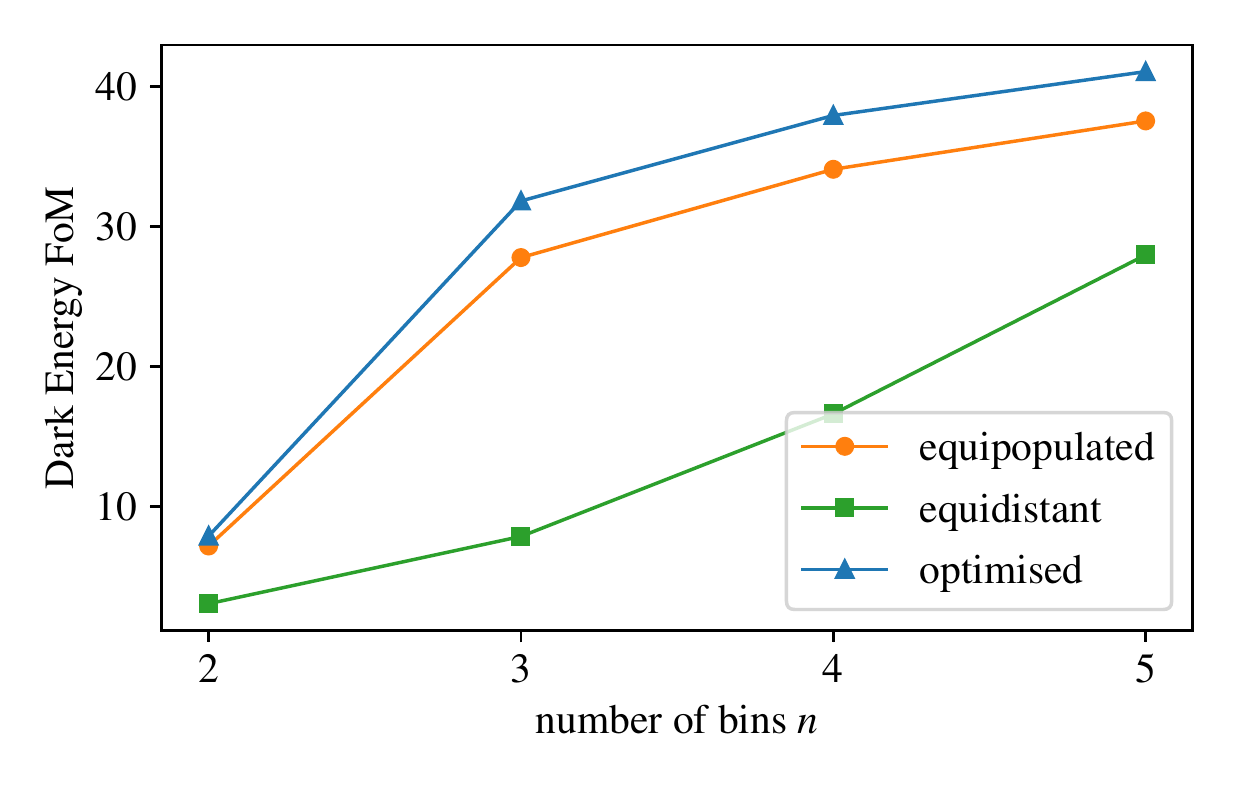}}
	\caption{\karl{Dark energy figure of merit as a function of the number of tomographic bins, for equipopulated binning (orange line, circles), for equidistant binning (green line, squares) and for optimised binning (blue line, triangles).}}
	\label{defom_scaling}
\end{figure}

\begin{figure*}
\resizebox{0.9\hsize}{!}{\includegraphics{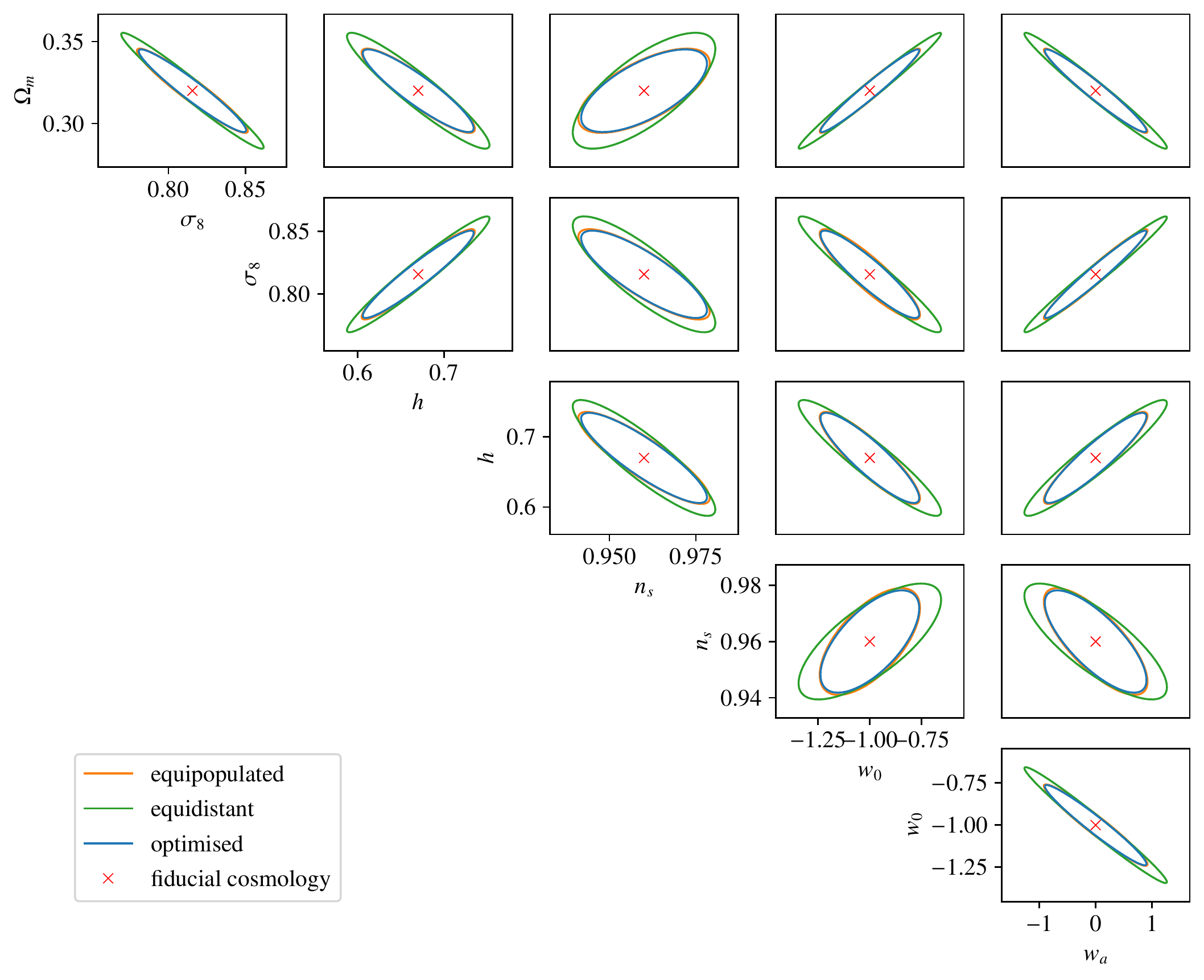}}
\caption{\karl{Error ellipses on a $w_0w_a$CDM parameter set with equipopulated binning (orange), equidistant binning (green) and optimised binning (blue), for 5 bin tomography. The optimisation target was the dark energy figure of merit. The fiducial cosmology is marked with a red cross.}}
\label{defom_ellipses}
\end{figure*}

{\karl{Of particular interest to the next generation of large-scale structure surveys is the dark energy figure of merit, which combines the (inverse) uncertainties on $w_0$ and $w_a$. As such, a survey yielding high values for the figure of merit are able to more precisely address the questions if dark energy has an equation of state equal to that of the cosmological constant today and if the equation of state has evolved or not. The scaling of the dark energy equation of state is shown in Fig.~\ref{defom_scaling}, suggesting that optimised binning surpasses conventional binning by as much as a factor $\sim 1.3$, but it needs to be pointed out that in this case one sacrifices precision in the remaining parameter set for obtaining small errors in $w_0$ and $w_a$: The figure of merit is computed from the marginalised errors which contain the uncertainties in the full parameter set, such that the improvement is rather generated by reducing the uncertainty in remaining parameters than reducing that in $w_0$ and $w_a$ directly. This effect can also be seen in the triangular plot in Fig.~\ref{defom_ellipses}, where the ellipse in the $w_0-w_\mathrm{a}$-plane and by that the dark energy figure of merit shrinks significantly. However, this can be counteracted by using a strong prior on other parameters, as for example provided by cosmic microwave background measurements. Furthermore, the degeneracy between $w_0$ and $w_\mathrm{a}$ gets slightly broken with the optimised redshift binning. This, again, is due to the fact that the bins are placed at higher redshift, such that the dependence on the scale factor of $w(a)$ becomes more pronounced in the analysis.}}

\begin{figure}
\resizebox{0.98\hsize}{!}{\includegraphics{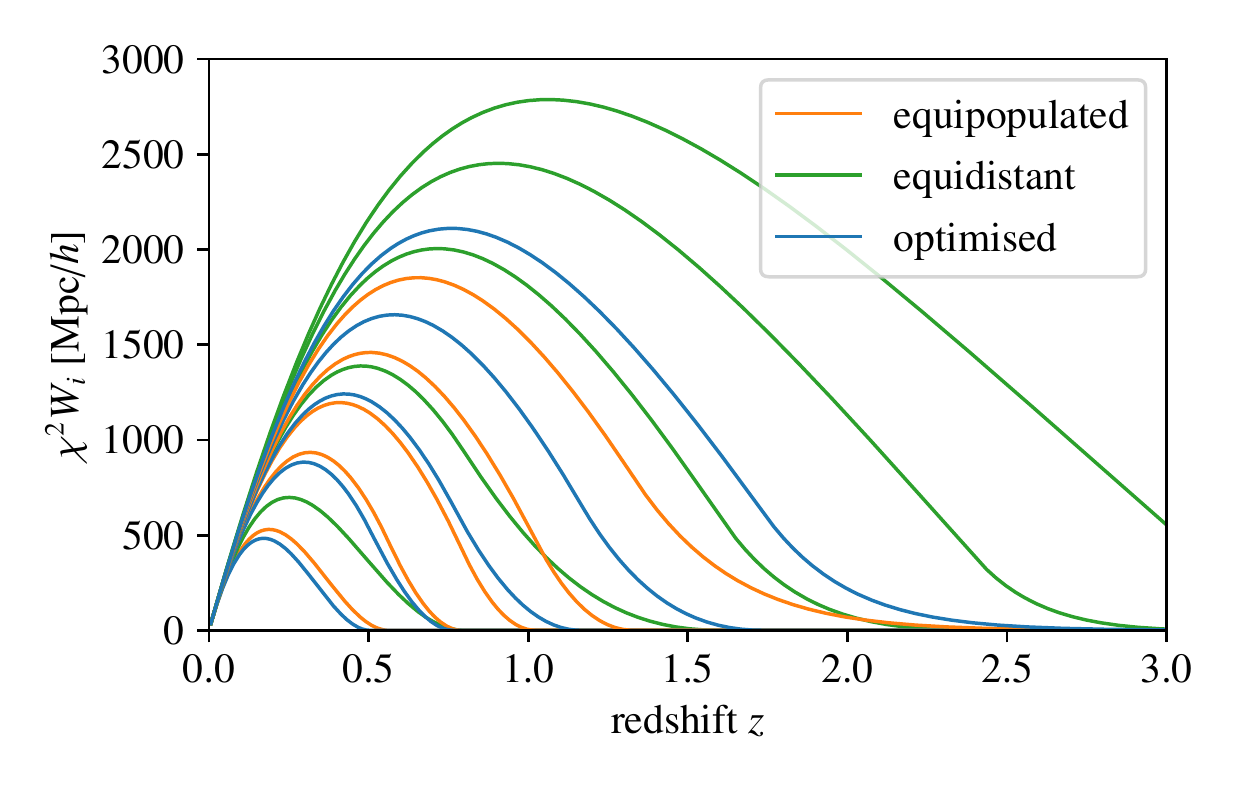}}
\caption{\karl{Lensing efficiency for equipopulated binning (orange), equidistant binning (green) and optimised binning (blue) with respect to the dark energy figure of merit, in all cases for 5 tomographic bins.}}
\label{defom_weighting}
\end{figure}

\begin{figure}
\resizebox{0.98\hsize}{!}{\includegraphics{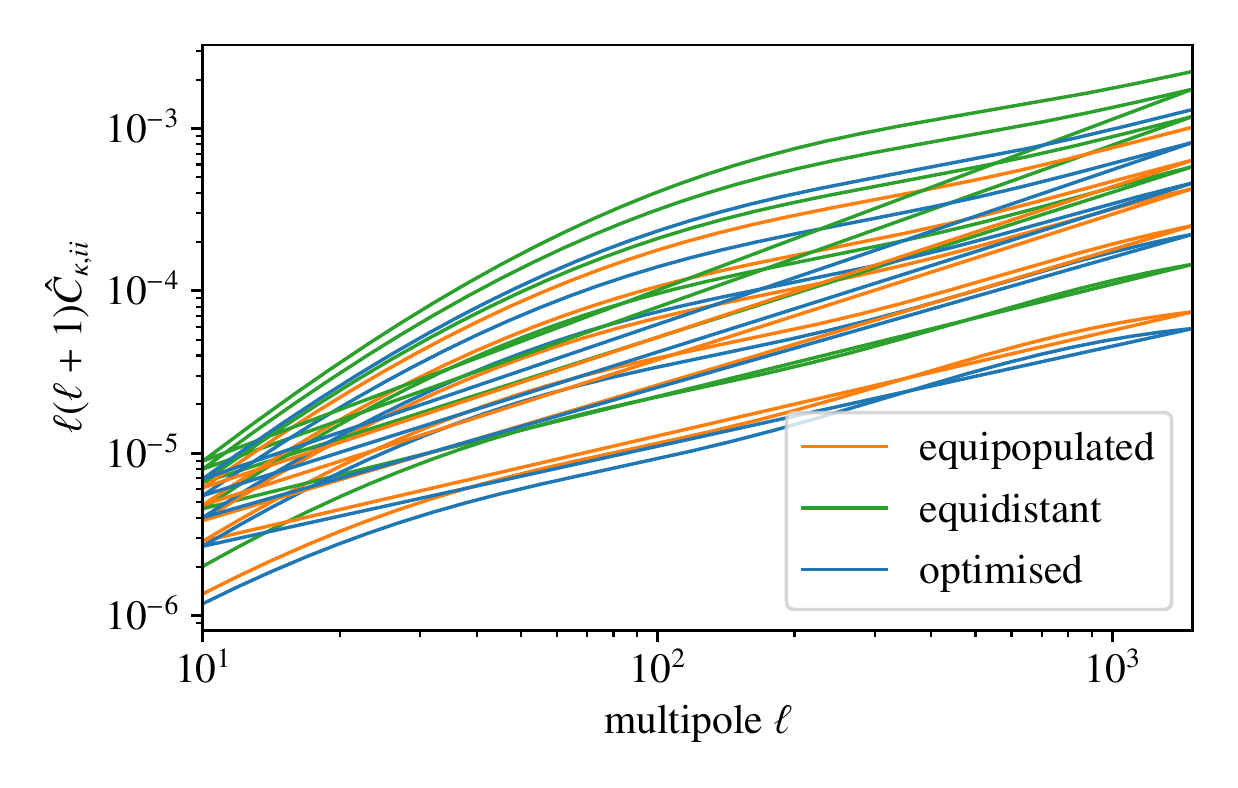}}
\caption{\karl{Spectra $C_{\kappa,ij}(\ell)$ of the weak lensing convergence for $i=j$ with the corresponding Poisson noise. The optimisation is computed for maximising the dark energy figure of merit.}}
\label{defom_spectra}
\end{figure}

{\karl{
For the same optimization target we show the corresponding weight functions and spectra in Fig.~\ref{defom_weighting} and Fig.~\ref{defom_spectra} respectively. From the first it can again, clearly be seen that the lensing weight is very close to the equipopulated case, but slightly enhanced. The former show this trend as well, however, also represents the enhancement of shot noise in some bins. This can be seen in the high-$\ell$ tail of the spectra. Clearly, the equidistant bins have a much stronger signal and weight in general, showing the trade of between more lensing but much higher shot noise.}}

\subsection{Optimisation of $\ln\det C$}
\begin{figure}
\resizebox{0.98\hsize}{!}{\includegraphics{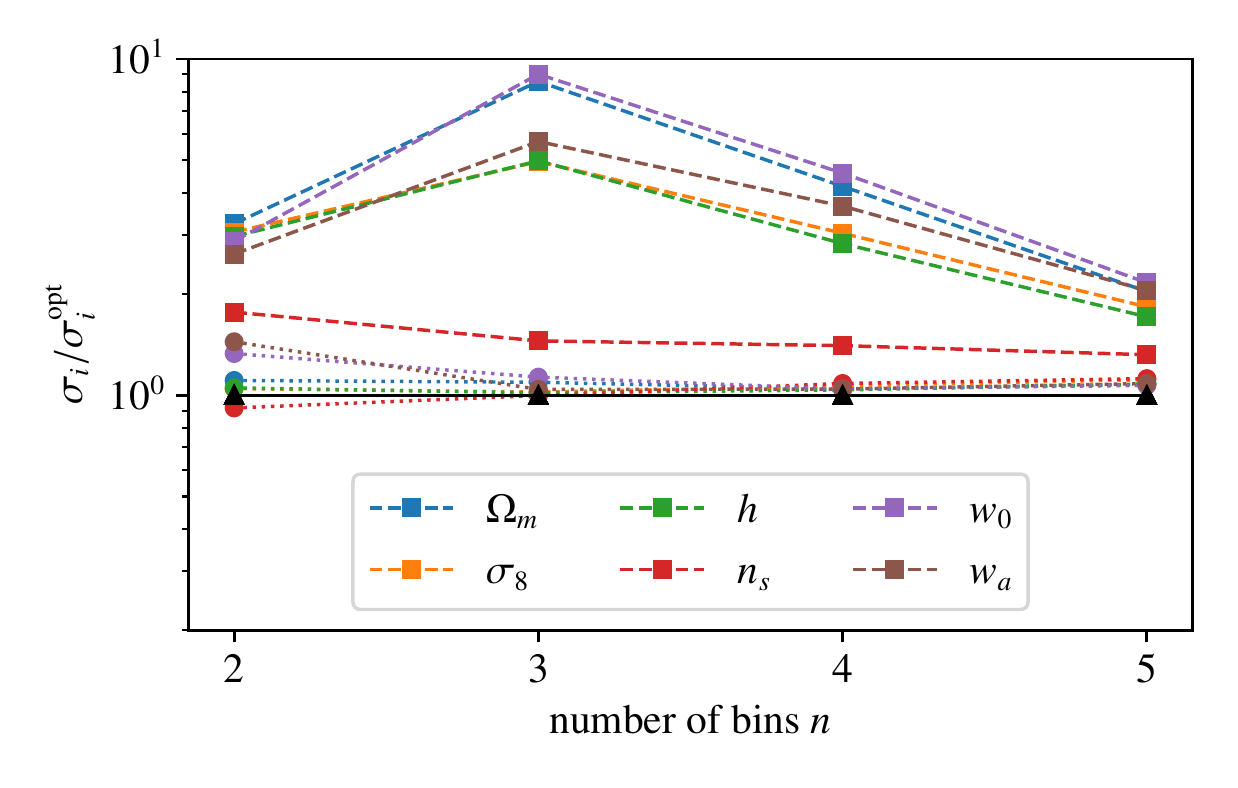}}
\caption{\spirou{Relative marginalised errors on parameters of a $w_0w_a$CDM-model, for equipopulated binnig (dotted lines, circles) and equidistant binning (dashed lines, squares), relative to an optimised binning (solid line, triangles) with the optimisation target $\ln \det C$.}}
\label{lndetc_errors}
\end{figure}

{\karl{
The optimisation of $\ln\det \boldsymbol{C}$ minimises the error budget with a relatively weak down-weighting of small errors, due to the convexity of the logarithm. In Fig.~\ref{lndetc_errors} one can see the result of this optimisation. Clearly, the optimisation yields improvements especially for $w_0$ and $w_a$, but as well for all other parameters. As discussed before $\ln\det C$ measures the volume of the parameter space of the decorrelated model parameters. {\spirou{As a consequence of the degeneracy in the $\Omega_m,\sigma_8$-plane typical for gravitational lensing, the price for a better constraint is paid by losing sensitivity in $\sigma_8$ (relative to $\Omega_m$).}} {\spirou{Since $\sigma_8$ is less correlated with the other parameters compared to $\Omega_m$ the optimisation sacrifices sensitivity in $\sigma_8$ first. Here we can see that the optimization yields the largest increases when very few bins are considered. It the omptimization has a considerable effect it is maximal at three redshift bins and decreases from there. For $n\to\infty$ the lines will approach unity since all the information in the field is recovered.}}}}

\subsection{Optimisation of total signal to noise-ratio}\label{sect_s2n}
Closely related to the computation of the Fisher-matrix is the cumulated signal to noise-ratio $\Sigma$,
\begin{equation}
\Sigma^2 = \sum_\ell \frac{2\ell+1}{2}\left(\hat{C}^{-1}_{\psi,ij}(\ell)S_{\psi,jk}(\ell)\:\hat{C}^{-1}_{\psi,kl}(\ell)S_{\psi,li}(\ell)\right),
\label{eqn_s2n}
\end{equation}
for a covariance $\hat{C}_{\psi,ij}(\ell) = S_{\psi,ij}(\ell) + N_{\psi,ij}(\ell)$ that splits into the signal $S_{\psi,ij}(\ell) = C_{\psi,ij}(\ell)$ and noise part $N_{\psi,ij}(\ell) = \frac{\sigma_\epsilon^2}{\ell^4}\frac{1}{\bar{n}f_i} \delta_{ij}$. It is equivalent to ask for the statistical error $\sigma_{\sigma_8}^2 = F^{-1}_{\sigma_8\sigma_8}$ of an unknown overall amplitude $\sigma_8$ of $S_{\psi,ij}(\ell)$ and to compute $\Sigma = \sigma_8/\sigma_{\sigma_8}$. As such, $P(\Sigma) = \mathrm{erf}(\Sigma/\sqrt{2})$ gives the cumulative probability that a correlation of the observed amplitude or higher is just a fluctuation of the noise.

Optimising $\Sigma$ yields a very modest improvement of a few percent over the standard binning, which is reflected by a curious, but not surprising coincidence that for this application the standard binning with equal fractions of the galaxy sample is already very close to the optimal one. {\spirou{One can find an analytical argument for this: For a pure amplitude the sensitivity is multipole and redshift independent and therefore just given by the pure signal compared to the noise which itself is entirely determined by the fiducial cosmology and the experimental setting, i.\,e. the number of galaxies and the number of redshift bins.}} In fact, setting up a simple model for investigating eqn.~(\ref{eqn_s2n}) with diagonal signal and noise covariances for every multipole, $C_i = S_i + N_i$ such that the trace relation becomes
\begin{equation}
\Sigma^2 = \sum_i \frac{S_i^2}{(S_i+N_i)^2} = \sum_i \frac{1}{(1+f_i)^2},
\end{equation}
with the proportionalities $S_i\propto 1/f_i^2$ and $N_i \propto 1/f_i$, the ideal binning can be obtained by variation with respect to the fractions $f_i$ subject to the constraint $\sum_i f_i = 1$. This can be incorporated by means of a Lagrange multiplier,
\begin{equation}
\Sigma^2(f_i,\lambda) = \sum_i \frac{1}{(1+f_i)^2} + \lambda\left(\sum_i f_i - 1\right).
\end{equation}
Then, the conditions $\partial\Sigma^2/\partial\lambda = 0$ and $\partial\Sigma^2/\partial f_j = 0$ imply in fact $f_j = 1/n$, with {\spirou{the Lagrange-multiplier being $\lambda = 2/(1+1/n)^3$.}}

This result is consistent with the case of optimising a measurement of $\sigma_8$ as the single parameter of a cosmological model, such that, unlike in the previous cases, uncertainties in other parameters do not enter that in $\sigma_8$ in the marginalisation process.

\subsection{Optimisation of the Kullback-Leibler-divergence $D_\mathrm{KL}$}
An alternative motivation for optimising tomographic binning could be to maximise the decrease in information entropy $\Delta S$ between the prior (for which we employ a CMB-prior) and the posterior one obtains after the weak lensing measurement. This information entropy $\Delta S$ would make sure that the reduction in total uncertainty, for instance expressed by the Cram{\'e}r-Rao-bound $\sigma_\mu^2\geq (F^{-1}_{\mu\mu})$, is maximised. The specific expression for the Kullback-Leibler-divergence is obtained by inserting the Fisher matrices into the general expression for for Gaussian likelihoods,
\begin{equation}
D_\mathrm{KL} = \frac{1}{2}\left(\mathrm{tr}\left((\boldsymbol{F}_\mathrm{CMB}+\tilde{\boldsymbol{F}})\boldsymbol{F}_\mathrm{CMB}^{-1}\right) - k + \ln\left(\frac{\det \boldsymbol{F}_\mathrm{CMB}}{\det(\boldsymbol{F}_\mathrm{CMB}+\tilde{\boldsymbol{F}})}\right)\right),
\end{equation}
with the number $k=4$ of cosmological parameters with marginalised weak lensing Fisher-matrix $\tilde{\boldsymbol{F}}$ and the CMB-prior $\boldsymbol{F}_\mathrm{CMB}$ for the remaining cosmological parameters $\Omega_m$, $\sigma_8$, $h$ and $n_s$.

\begin{figure}
\resizebox{0.98\hsize}{!}{\includegraphics{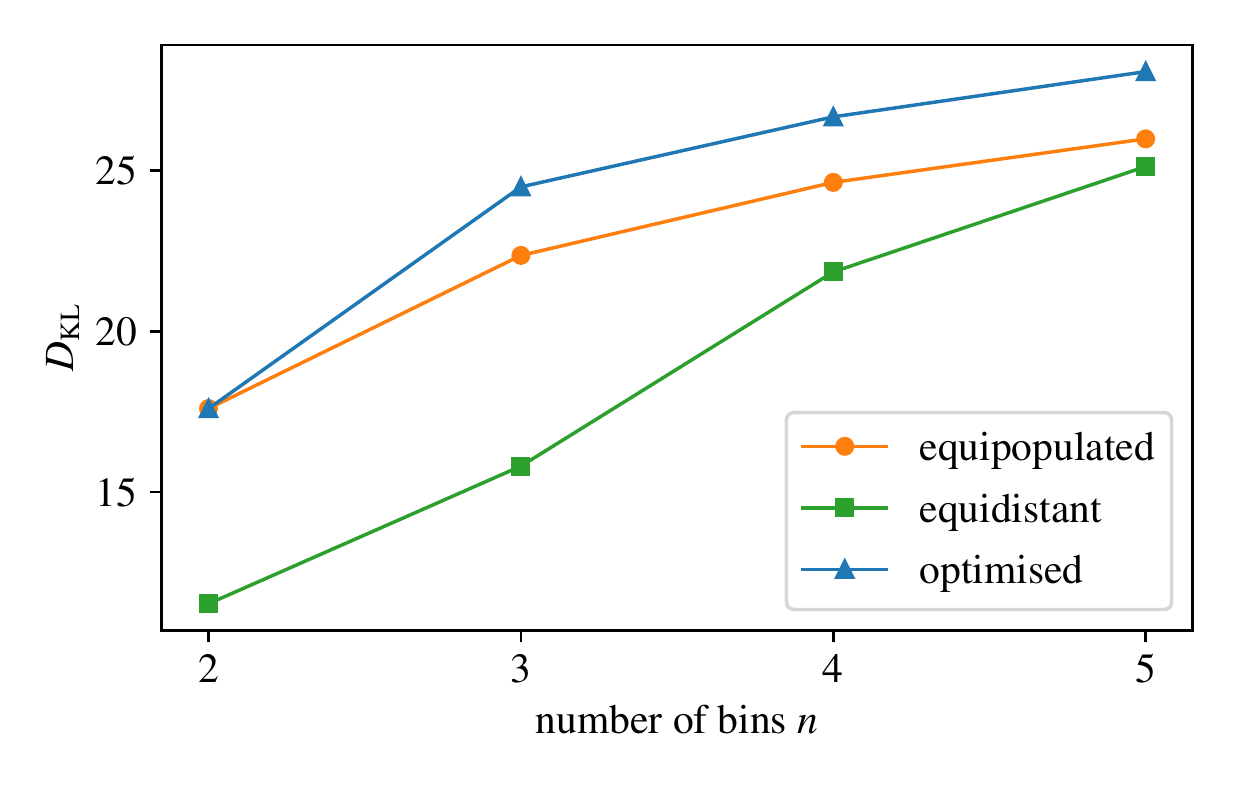}}
\caption{\karl{Difference in information entropy as a function of the number of tomographic bins, for equipopulated binning (orange line, circles), equidistant binning (green line, squared) and for optimised binning (blue line, triangles) to yield the largest relative information entropy. $D_\text{KL}$ is given in units of nats.}}
\label{kullback_scaling}
\end{figure}

{\spirou{Fig.~\ref{kullback_scaling} shows how the information entropy decrease (and therefore the information gain) between a CMB-prior and the weak lensing measurement would scale with the number of tomographic bins. There is a steady change in information entropy as expressed by the Kullback-Leibler divergence as a function of the number of tomographic bins, where optimised and equipopulated binning show a very similar behaviour. The equidistant binning falls behind by a large margin and only catches up with a larger number of tomographic bins. However, the relative gain of the optimal binning strategy with respect to a CMB prior is only between one and two percent. This is consistent with the observations made in Figs.~\ref{defom_ellipses} and \ref{lndetc_errors} showing that the overall errors and therefore information gain does not change significantly. However, the sensitivity on individual target parameters can be enhanced strongly, i.e. for applications where the conditionalised error matters.}}

\begin{figure*}
\resizebox{0.98\hsize}{!}{\includegraphics{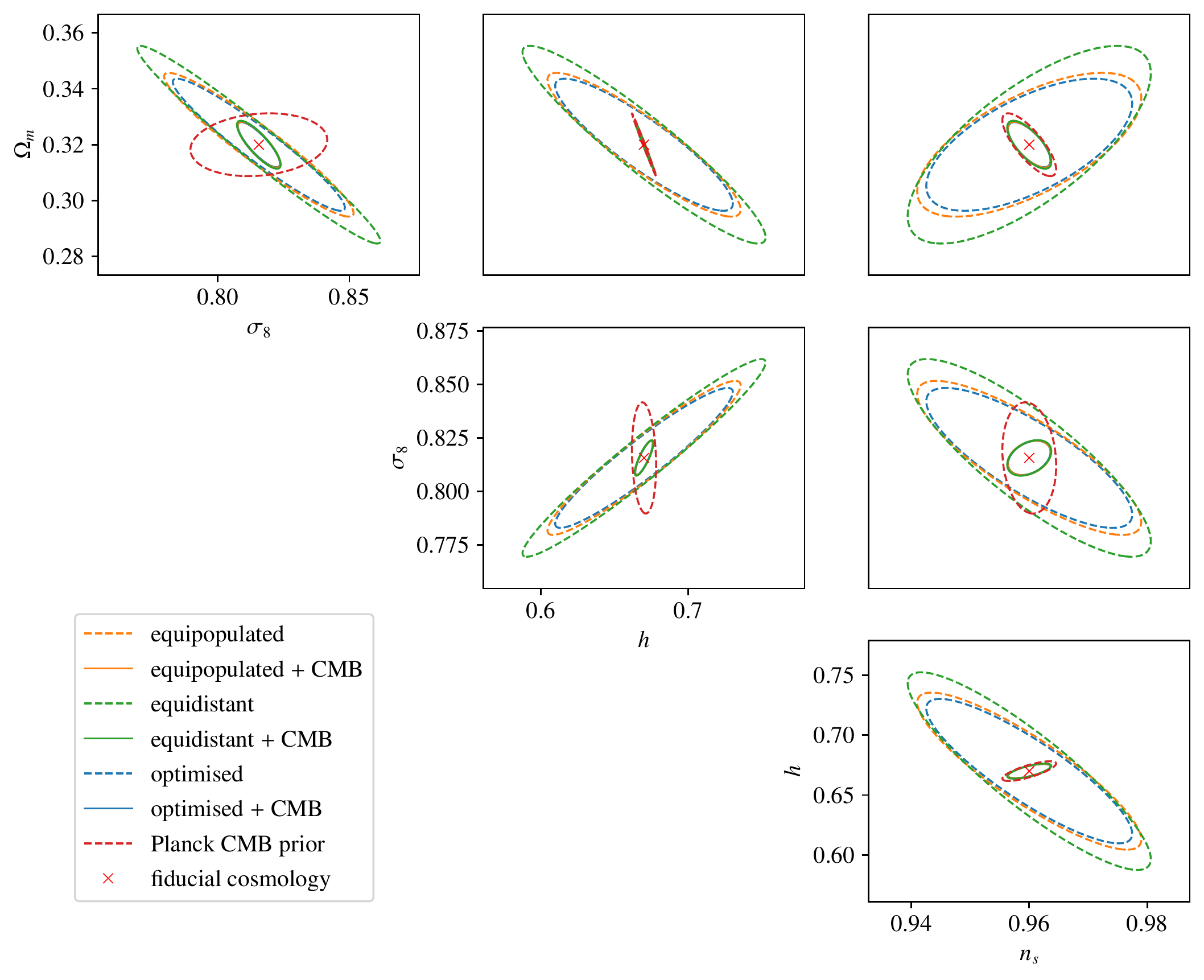}}
\caption{\spirou{Error ellipses for equipopulated binning (orange lines), equidistant binning (green lines) and optimised binning (blue lines), for 5-bin tomography, along with a Planck CMB-prior (dashed lines, red). The optimisation target was maximising the Kullback--Leibler-divergence $D_\mathrm{KL}$ between prior and posterior, and the line styles differentiate between the information gain of a weak lensing experiment over the CMB alone and that of a combined weak lensing and CMB-likelihood, in our cases \textsc{Euclid} and \textsc{Planck}, over the CMB alone.}}
\label{kullback_errors}
\end{figure*}

The corresponding error ellipses in the parameters $\Omega_m$, $\sigma_8$, $h$ and $n_s$ are illustrated in Fig.~\ref{kullback_errors}, for the case of 5 tomographic bins, where the previous plot already showed that only marginal gains can be expected. Consistent with this observation are very similar error ellipses for the combined measurements.

We would expect that generalisations to the information entropy should yield comparable results, for instance if the Shannon-entropy, which we essentially use here, because it enables analytical results for Gaussian distributions, is replaced by R{\'e}nyi entropies. In a larger context, we would argue that bin optimisations on the basis of (Bayesian) evidence could help to differentiate between competing cosmological models, too. In this application, one could formulate the Bayesian evidence ratio between two models for a given prior as a function of the tomography bin edges and determine the optimised binning. While this would be straightforward for Gaussian likelihoods and priors, non-Gaussian cases would need MCMC-evaluated evidences, which comes at a higher computational cost compared to the Fisher formalism.

\section{Summary}\label{sect_summary}
Subject of this paper are optimisations of the tomographic bins for weak lensing: By introducing tomographic bins of varying width we aim to increase the sensitivity of weak lensing measurements and to reduce the statistical error on a $\Lambda$CDM and $w_0w_a$CDM parameter set. For this purpose, we {\spirou{work in a Gaussian approximation of the likelihood,}} derive measures for the total statistical uncertainty from the Fisher-matrix and vary the bin edges in redshift with a Nelder-Mead algorithm to yield the best possible result. We carried out numerical optimisation for $(i)$ individual marginalised cosmological parameters, $(ii)$ the trace of the parameter covariance, $(iii)$ the Frobenius-norm of the parameter covariance, $(iv)$ the determinant of the parameter covariance, which corresponds to the volume of the $1\sigma$-ellipsoid, $(v)$ the dark energy figure of merit and lastly, $(vi)$ the Kullback-Leibler-divergence relative to a Gaussian CMB-prior. In all cases, the Nelder-Mead algorithm performed well with a standard binning containing equal fractions of the galaxy population as the initial condition and for a standard choice of settings for the algorithm. {\karl{Depending on the target for optimisation we could demonstrate a slight reduction of the total and of individual statistical errors, but also that the equipopulated choice is very close to the optimal binning.}} {\spirou{Our specific application is \textsc{Euclid}'s weak lensing survey in conjunction with CMB-observations by \textsc{Planck}.}}

{\spirou{Our results show some differences to those found in \citet{2019arXiv190106495K} as we investigate the influence of different metrics on the optimisation and stay in redshift space for the optimisation. In particular we do not find that equally space redshift bins are always the best configuration, but would like to emphasise that we do not consider catastrophic outliers in the redshift estimation or systematics in the redshift assignments, but rather focus on different optimisation targets and the minimisation of statistical error. Our result is, that a relatively low number of bins with a finer binning at high redshift, places tighter requirements on photometry, but seems to be well below percent-errors scaling with $1+z$, as commonly quoted in the context of weak lensing \citep{abdalla_photometric_2008}. Overall, however, we find that equipopulated bins are not very far from the optimum for many different targets, which is reflected by a simplified analytical argument about optimisation of the signal to noise-ratio, where equipopulated binning can be shown under certain assuptions to be optimal.}}

{\spirou{The results for an optimisation of a Gaussian likelihood can not be transferred to the non-Gaussian case in a direct way, as the improvement is very weak. This is because one would need to optimise the proper error measure obtained by MCMC methods. These are, of course, more difficult to interpret and measures of total error, in particular information entropies such as the Kullback-Leibler divergence, and numerically challenging to obtain.}} The generalisation is, however, conceptually straightforward and would also yield similar improvements although and substantially higher computational costs. One way around this problem would be to use higher order approximation schemes such as the DALI expansion \citep{Sellentinetal, 2015arXiv150605356S}. Going beyond a straightforward optimisation of the likelihood, we successfully looked into information entropy measures and plan to consider optimisations on the basis of Bayesian evidence, in order to maximise the distinguishability between models {\spirou{and to sharpen the capability of weak lensing in differentiating between cosmological models.}}

We conclude by emphasising again that the optimisation shows that the constraints in the case of optimal binning saturate at a lower number of bins. This suggests that for \textsc{Euclid} a low number, typically $n_\mathrm{bin} = 4\ldots5$ of optimally chosen redshift bins is sufficient to extract the needed cosmological information thus rendering the redshift uncertainties much less troublesome than for $n_\mathrm{bin}\approx 10$ or more. Surely there is room for alternative binning strategies as discussed in \citet{taylor_preparing_2018} with implications for experimental design \citep{amara_optimal_2007}, in particular if systematic errors become important \citep{cardone_power_2014}.

\section*{Acknowledgements}
RR acknowledges funding through the HeiKA research bridge of the Karlsruhe Institute for Technology and Heidelberg University, as well as support by the Israel Science Foundation (grant no. 1395/16 and 255/18). BMS would like to thank the Universidad del Valle in Cali, Colombia, for their hospitality.

\bibliographystyle{mnras}
\bibliography{references}

\appendix
\section{Convergence}
\label{app:covergence}
{\spirou{The crucial aspect of the work is the optimisation procedure. In order to make sure that the Nelder-Mead algorithm finds a global minimum and not a local one two precautionary measures are taken. $(i)$ We run the optimisation from different starting points in redshift space and check for convergence; $(ii)$ the Nelder-Mead-algorithm resets the simplex size after it has shrunk below a certain threshold for the first time and walks through the optimisation surfaces with multiple cycles, additionally minimising the risk of finding a local minima instead of the global one. In \autoref{fig:convergence} in the top panel the resetting can clearly be seen: After a few cycles, the simplex is reset with a temporary deterioration of the optimsation target, which the recovers in the subsequent steps. The bottom panel directly shows the target function as a function of the bin edge redshifts. The simplex of the Nelder-Mead algorithm is depicted by the coloured lines as it evolves over the iterations. Again, one can see clearly the resetting as well as that the algorithm indeed finds the global minimum, which is indicated by the deepest blue in the colour shading. Furthermore, the equipopulated case is very close to the optimal one as we found before.}

\begin{figure}
	\resizebox{0.98\hsize}{!}{\includegraphics{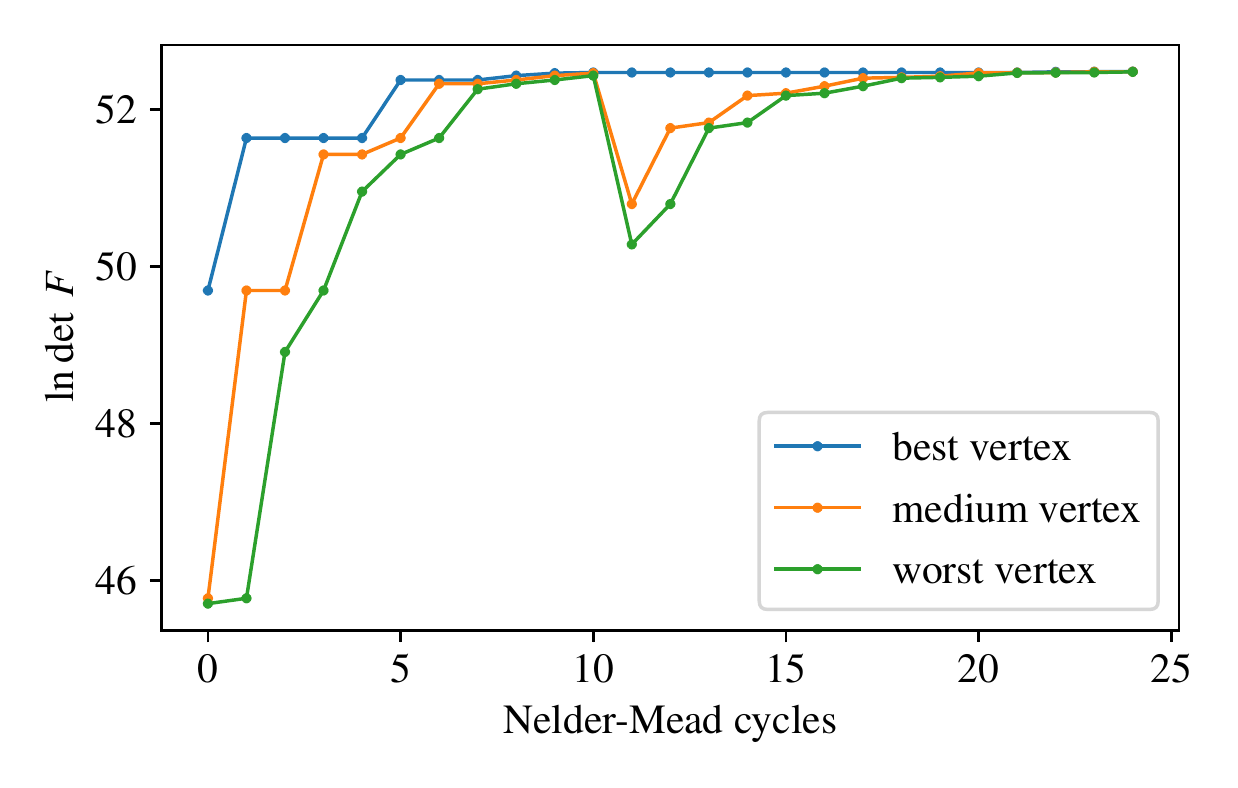}}
	\resizebox{0.98\hsize}{!}{\includegraphics{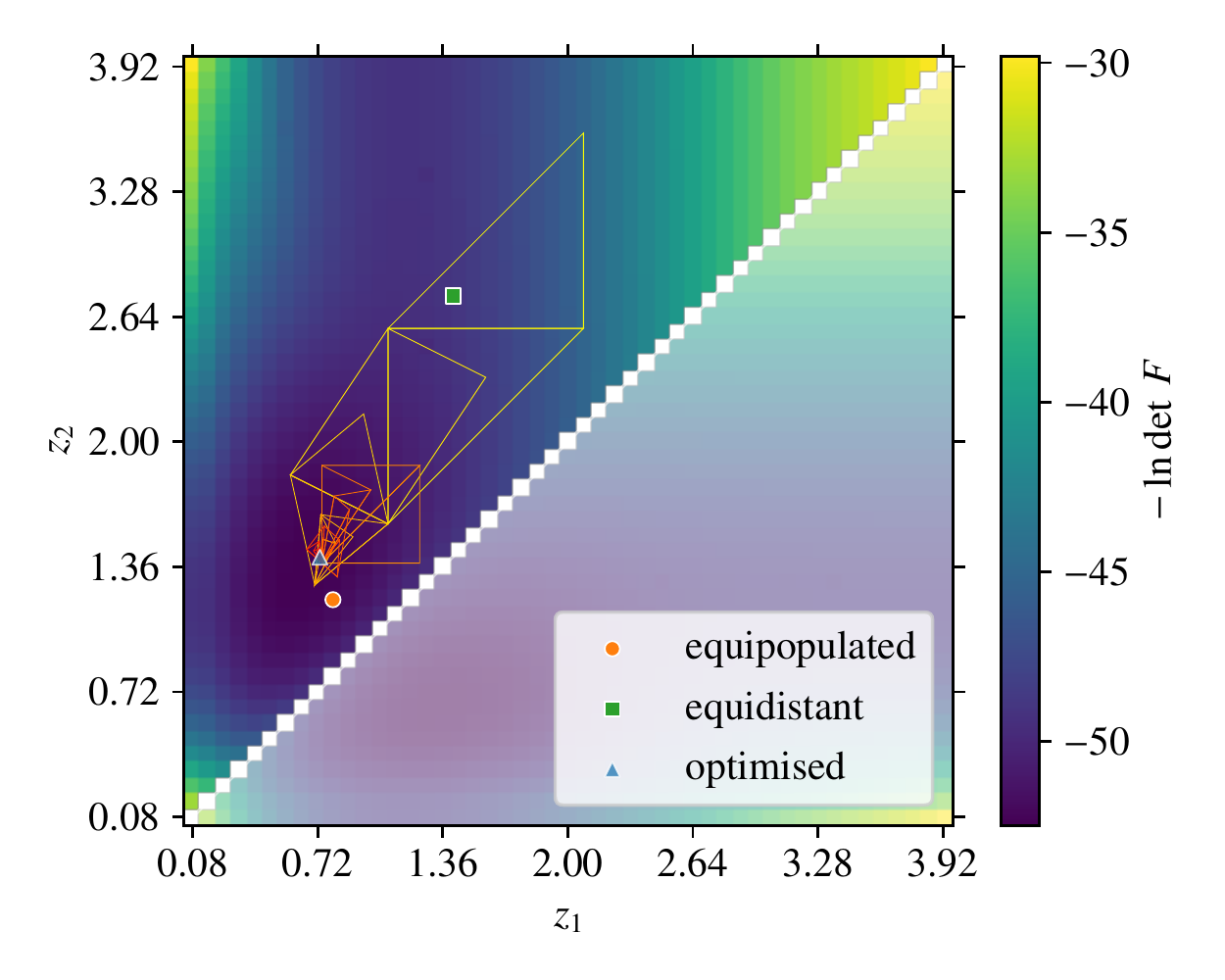}}
	\caption{\karl{Convergence of the Nelder-Mead algorithm for the logarithm of the figure of merit for three tomographic bins. \textit{Top}: The target function as a function of cycles for different vertices. The algorithm resets its simplex occasionally to minimise the risk of being stuck in a local minimum. \textit{Bottom}: Same scenario as in the top panel but now the target function is directly shown as a function of two bin edge redshifts in the course of the optimisation, with regular resettings. It can be seen that the equipopulated case is very close to the global minimum.}}
	\label{fig:convergence}
\end{figure}

\bsp
\label{lastpage}
\end{document}